\documentclass[12pt]{article}
\usepackage{epsfig}

\begin{document}

\title{Influence of the Cut-off Elevation Angle and Elevation-Dependent
Weighting on Parameter Estimates: \\ A Case of CONT05}
\author{Zinovy Malkin}
\date{Pulkovo Observatory, St. Petersburg, Russia}
\maketitle

\begin{abstract}
In this paper, results are presented on studies which have been performed
to investigate the impact of the cut-off elevation angle (CEA)
and elevation-dependent weighting (EDW) on the EOP estimates and baseline
length  repeatability.
For this test, CONT05 observations were processed with different CEA
and EDW, keeping all other options the same as used during the routine
processing. Uncertainties and biases, as well as
correlations between estimated parameters have been investigated.
It has been shown that small CEA, up to about 8--10 degrees does not
have large impact on the results, and applying EDW allows us to get
better result (smaller errors). However, this result has been proven
with standard geodetic VLBI observations, where rather few observations
were made at low elevations. Perhaps, special R\&D sessions with more
uniform distribution of observations over elevation may be useful for
more detailed study on the subject.
\end{abstract}

\vfill
\noindent \hrule width 0.4\textwidth
~\vskip 0.2ex
\noindent {\small 5th IVS General Meeting, St.~Petersburg, Russia, 3--6 March 2007}
\eject

\section{Introduction}

It is well known that precision and accuracy of astronomical observations,
both optical and radio,
made through the Earth's atmosphere depend on the elevation at which the
object is observed.
These errors grow with decreasing of the elevation due to larger air
mass and difficulties in modelling of refraction effects at low elevation.
From this point of view observations should be made in the near-zenith
zone when possible.

On the other hand, inclusion in processing of observations made at
low elevations is important when definite groups of highly correlated
parameters, for instance station coordinates and zenith troposphere
delays, are estimated simultaneously.  In such a case using observations
made at in a widest range of elevation allows us to mitigate the
correlations between unknowns and improve the solution.

To meet these mutually exclusive requirements, proper elevation-dependent
weighting (EDW) of observations is used.
In a special case of step-like weighting function, i.e. rejection of
the observations made at the elevation less than the given limit,
such a limit usually is called cut-off elevation angle (CEA).

It was shown in many studies that elevation-dependent weighting may have
a significant impact on the results of processing of the space geodesy
observations.  In particular, several studies of this effect was made
by the Goddard and Vienna VLBI analysis groups
in the framework of the IVS VLBI2010 Committee
activity\footnote{href="http://ivscc.gsfc.nasa.gov/pipermail/ivs-v2c/}.
They investigated an influence of CEA and EDW
on geodetic results such as Earth orientation parameters (EOP),
baseline length repeatability, troposphere parameters,
and station heights.  Those results were based on simulation.
Gipson in \cite{Gipson07} used another approach to EDW.
He applied elevation dependent additive noise to the measurement error
instead of using a weighting factor as it is usually being made.
He tested his method with the actual CONT05 observations.
Results of both mentioned and other results are sometimes contradictory.
This gave an impulse to the present work where
some results are presented of investigation of the impact of the CEA
and EDW on the baseline length repeatability and EOP estimates.

\section{Test description}

For this test, CONT05A observations were processed making use of
OCCAM software with different EDW functions including CEA,
keeping all other options as follows:
\begin{itemize}
\itemsep=-0.3ex
\item Kalman filter mode (KF),
\item random walk model for clocks, PSD=1.5 ps$^2$/s,
\item random walk model for ZTD, PSD=0.25 ps$^2$/s,
\item one NS and EW troposphere gradient estimate for the session.
\end{itemize}

\begin{figure}
\centering
\epsfclipon \epsfxsize=0.48\textwidth \epsffile{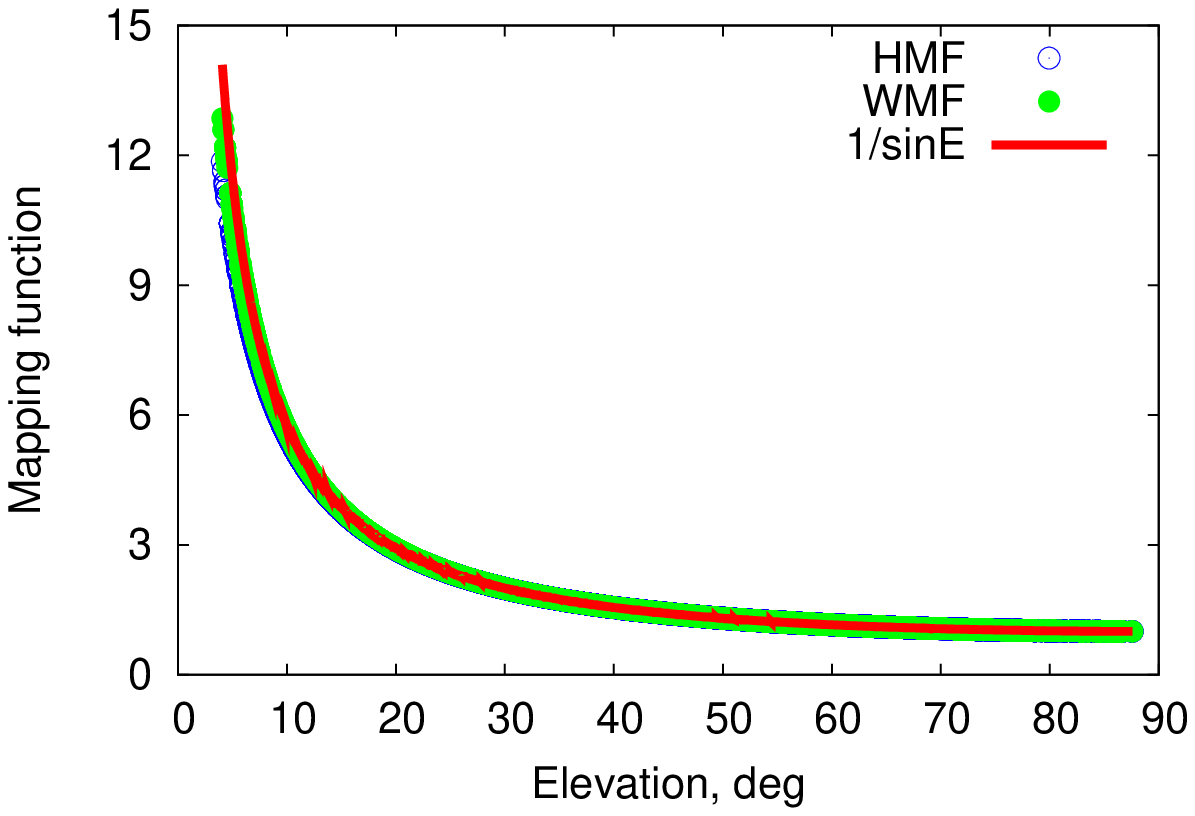}
\hspace{0.02\textwidth}
\epsfclipon \epsfxsize=0.48\textwidth \epsffile{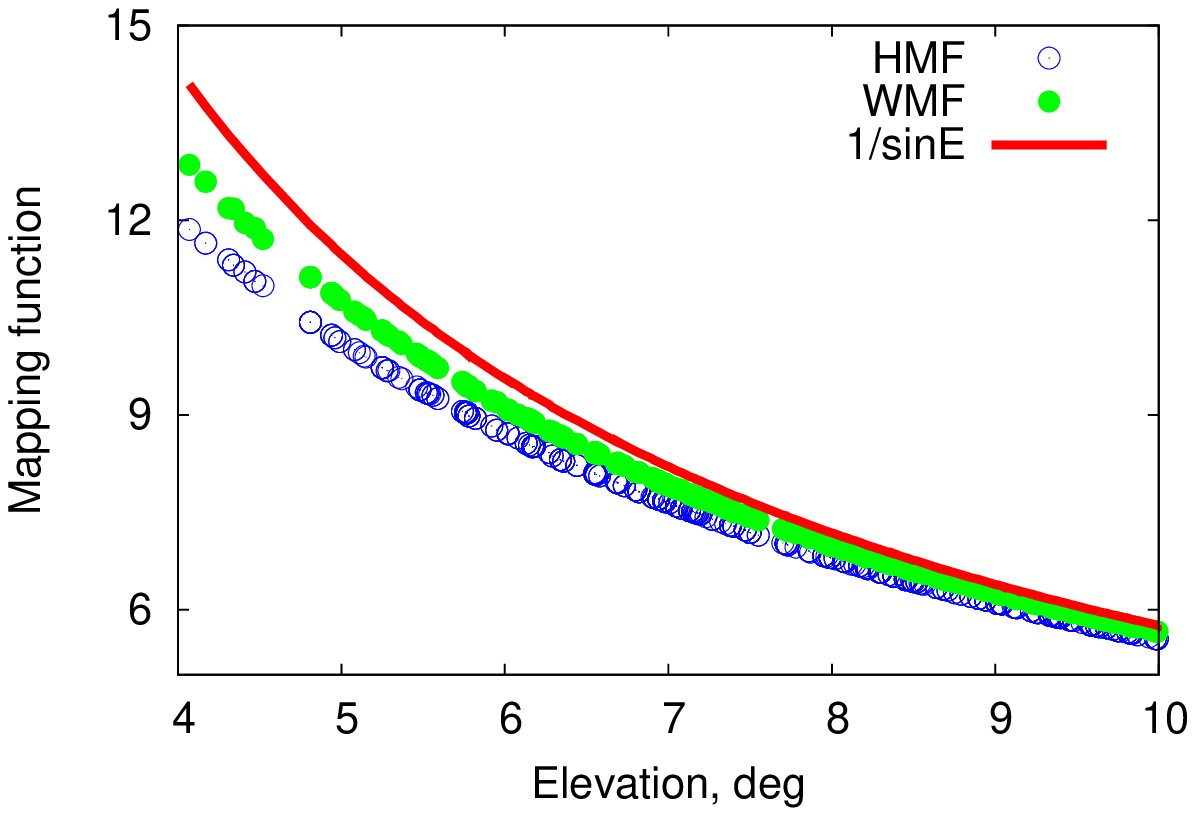}
\caption{Actual values of the hydrostatic (HMF), wet (WMF) mapping functions
for the C0509 (05SEP20XA) session (one point correspond to one observation).
Solid line corresponds to function $1/\sin e$.
Data for the full elevation range ({\it left}) and zoomed data for
low elevations ({\it right}) are shown.}
\label{fig:mf}
\end{figure}

For the continuous EDW mode (continuous weighting function),
the measurement error is multiplied by a factor
\begin{equation}
W_e=(\sin e_0\, / \sin e)^p \,,
\label{eq:edw}
\end{equation}
where $e_0$ and $p$ are EDW parameters,
$e$ is the source elevation.  Such a weighting function provides a smooth
stepless change in weight for any $e_0$.
One can see that a case of $e_0=90^\circ, \, p=1$ gives merely $W_e=1/\sin e$,
which is close to actual wet mapping function used in the last works by
MacMillan and Gipson (private communications).
Figure \ref{fig:mf} shows actual hydrostatic and wet mapping functions
along with approximation
$1/\sin e$ for a typical CONT05 session. One can see that these functions
are close enough, and either of them can be used for the EDW
without significant impact on the result.

EDW mode with $e_0=10, \, p=2$ was implemented in the OCCAM/GROSS
software \cite{OCCAM} for routine data processing.
It will be referred hereafter as "normal mode".

In a case of CEA we have
$$W_e =\left\{
\begin{array}{ll} 1, & \mbox{if $e>=e_0$} \\ 10^3 & \mbox{otherwise} \end{array}
\right.$$
The latter line corresponds to the KF realization used in OCCAM.

For VLBI delay, measurement error coming from correlator is multiplied
by two $W_e$ values computed for both the stations.
In our test, $e_0=3(2)25^\circ$ were used for CEA test, and
$e_0=10, 25, 45, 90^\circ, \, p=1, 2$ were used for continuous EDW mode.

\section{Test Results}

\subsection{Baseline length repeatability}

\begin{figure}
\centering
\epsfclipon \epsfxsize=0.48\textwidth \epsffile{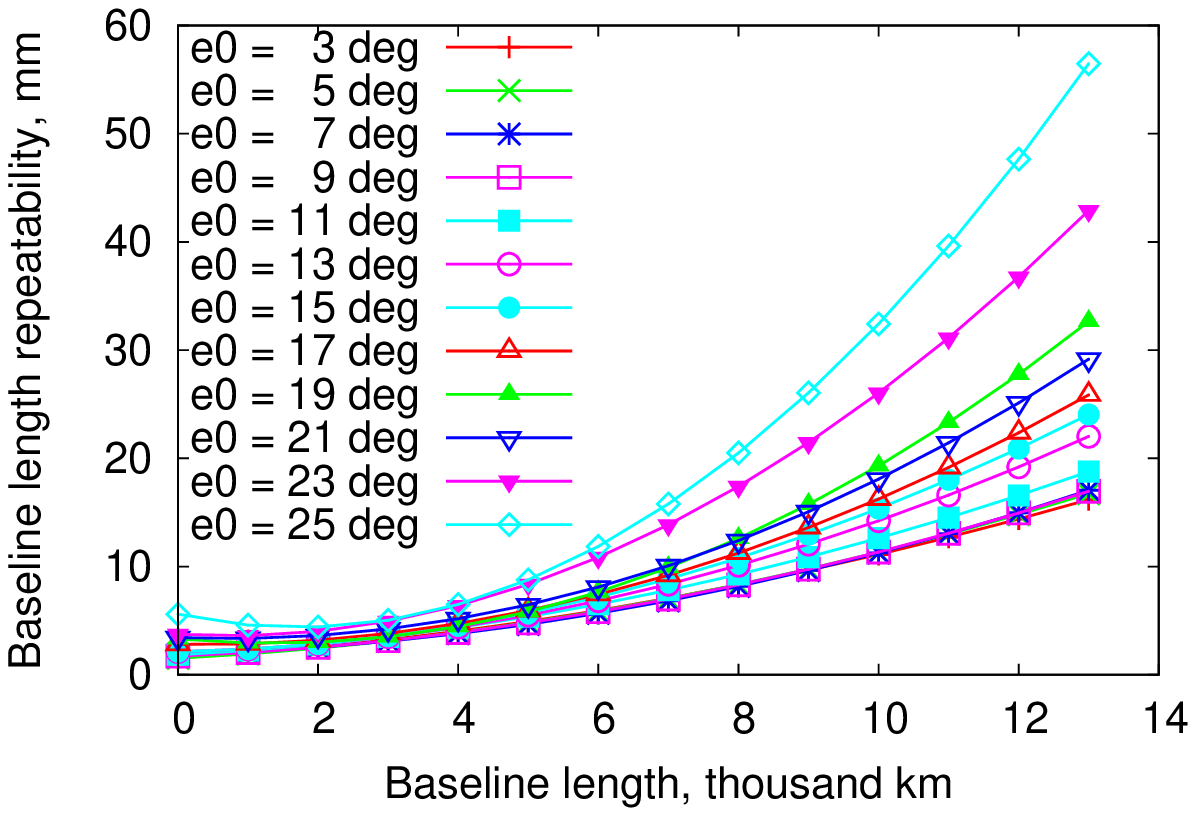}
\hspace{0.02\textwidth}
 \\ 
\epsfclipon \epsfxsize=0.48\textwidth \epsffile{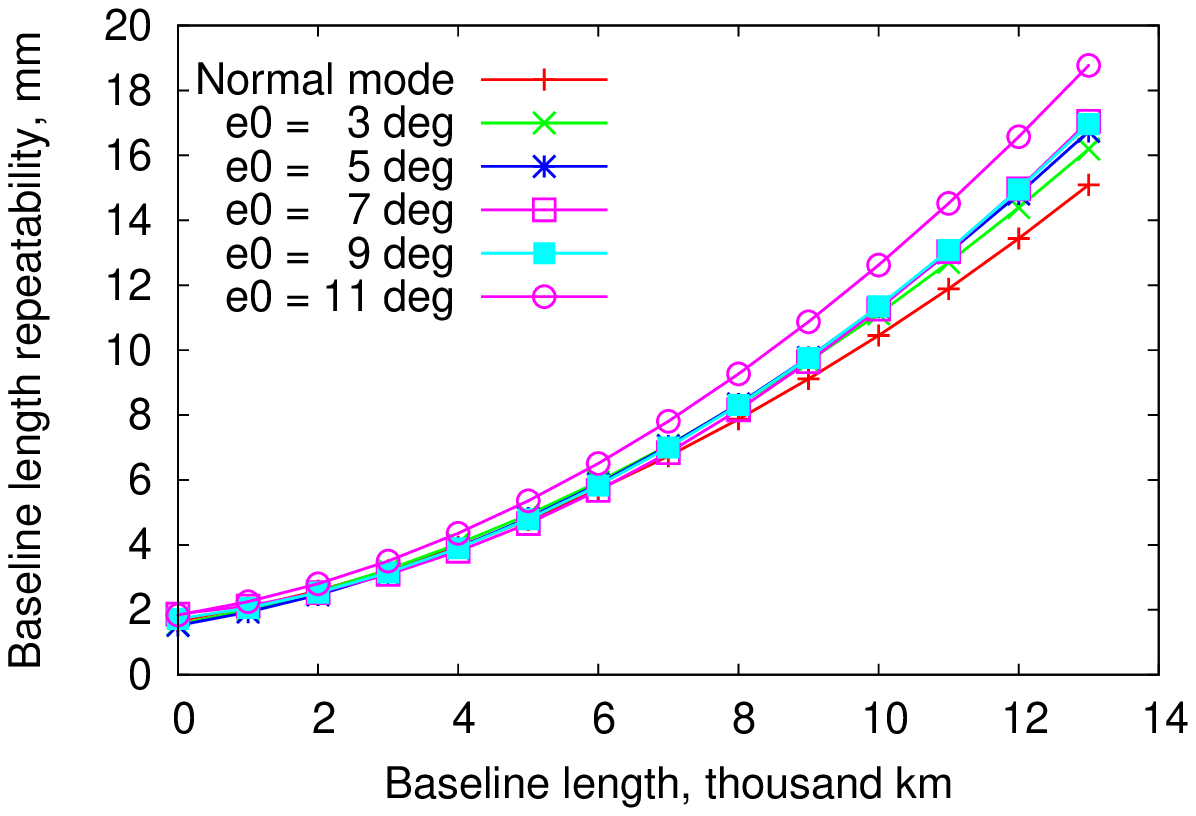}
\caption{Dependence of the baseline length repeatability on the cut-off
elevation angle (quadratic regression):  all tested
modes ({\it left}) and data for low elevations ({\it right}).}
\label{fig:cea_baselines}
\end{figure}

Test results obtained with different CEA are shown in
Figure \ref{fig:cea_baselines}.
The case of $e_0=3^\circ$ includes all the observations without
weighting, since no CONT05 observations were made at the elevation
less than 4$^\circ$.

Table \ref{tab:edw_baseline} shows EDW test results. Different EDW modes
are denoted as w\_e\_p, where e and p are $e_0$ and $p$ in Eq.(\ref{eq:edw}).
Test results are given for quadratic approximation in percent
with respect to the case of CEA with $e_0=3^\circ$.
One can see that several EDW modes show about the same
improvement in the baseline length repeatability.

\begin{table}
\centering
\caption{Comparison of the baseline repeatability obtained with
different EDW models. See explanation in text.}
\label{tab:edw_baseline}
\begin{tabular}{|c|c|c|c|c|}
\hline
EDW mode & \multicolumn{4}{|c|}{Baseline length, $10^3$ km} \\
\cline{2-5}
         &   3   &   6   &   9   &   12  \\
\hline
w\_10\_1 & ~98.0 & ~97.2 & ~97.3 & ~97.6 \\
w\_10\_2 & ~95.3 & ~95.0 & ~95.5 & ~96.1 \\
w\_25\_1 & 102.6 & ~92.8 & ~87.2 & ~83.8 \\
w\_25\_2 & ~95.7 & ~91.8 & ~91.4 & ~91.9 \\
w\_45\_1 & 101.7 & ~90.0 & ~84.6 & ~81.9 \\
w\_45\_2 & 104.0 & ~97.2 & 101.8 & 107.6 \\
w\_90\_1 & 101.3 & ~90.5 & ~85.4 & ~82.8 \\
w\_90\_2 & 109.6 & 127.3 & 161.1 & 191.1 \\
\hline
\end{tabular}
\end{table}

\subsection{EOP}

\begin{figure}
\centering
\epsfclipon \epsfxsize=0.48\textwidth \epsffile{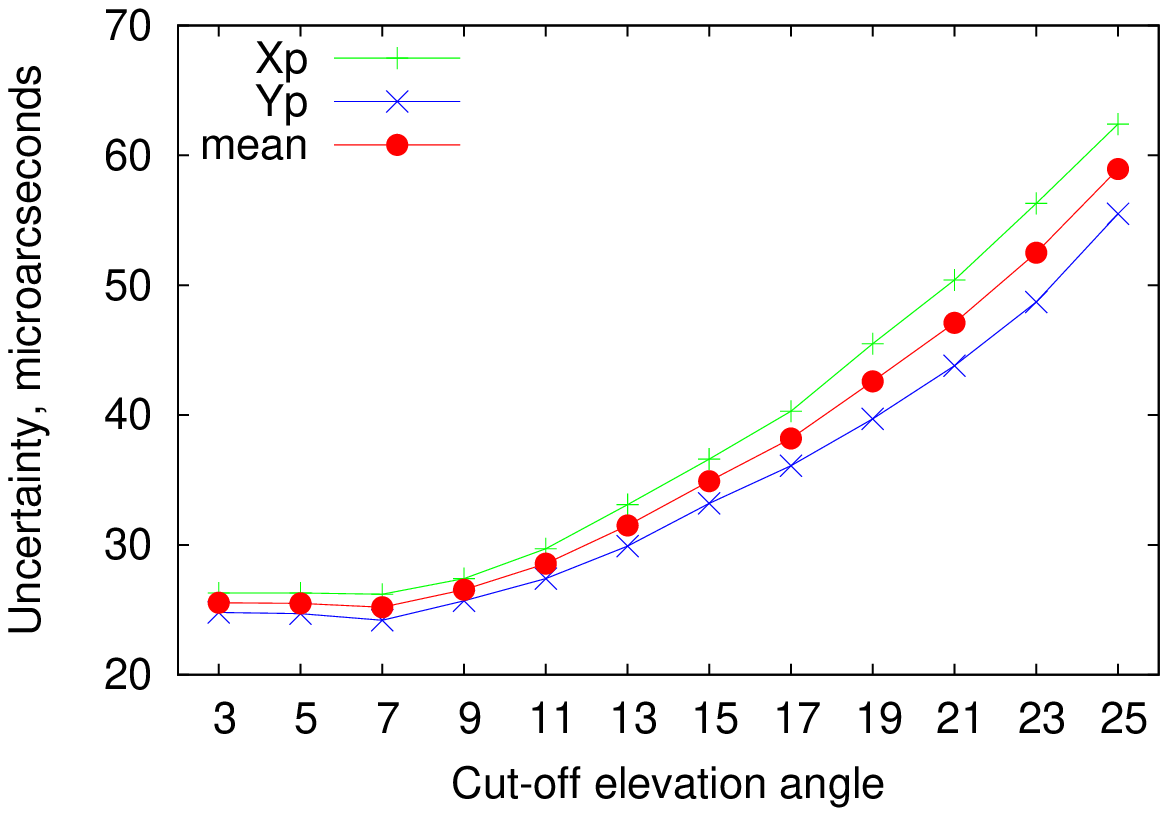}
\hspace{0.02\textwidth}
\epsfclipon \epsfxsize=0.48\textwidth \epsffile{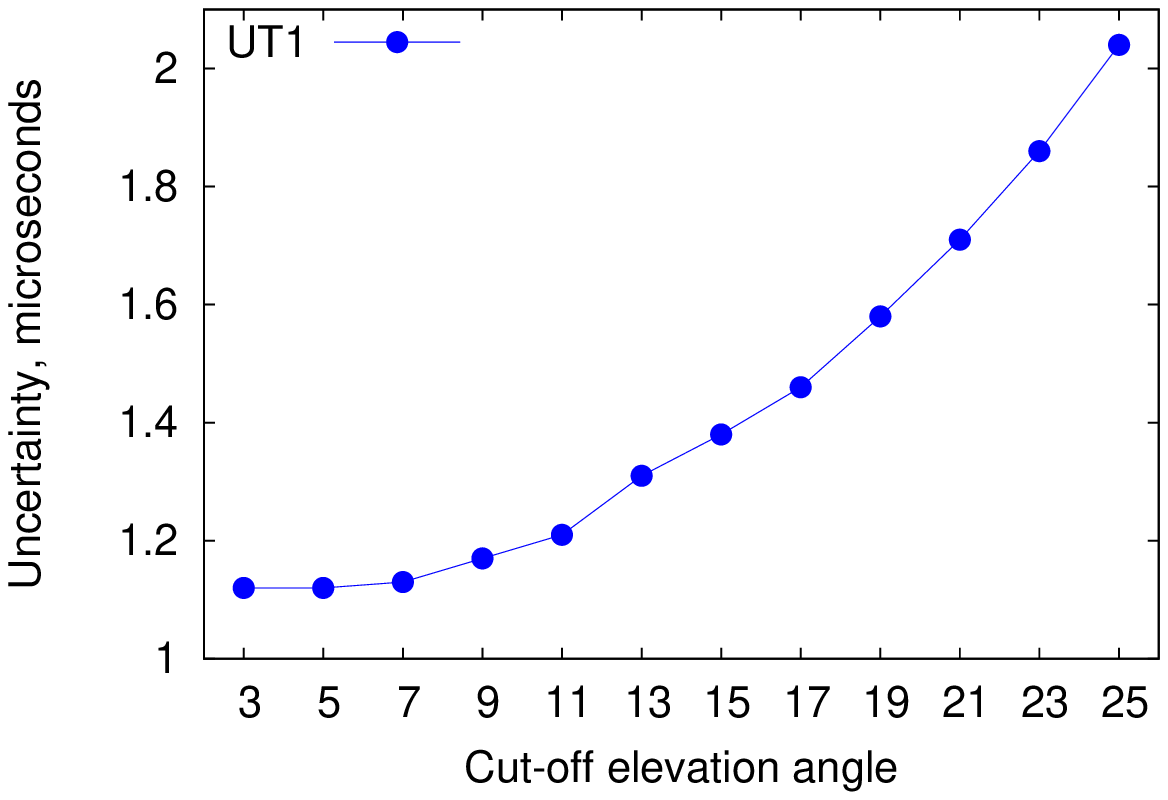} \\
\epsfclipon \epsfxsize=0.48\textwidth \epsffile{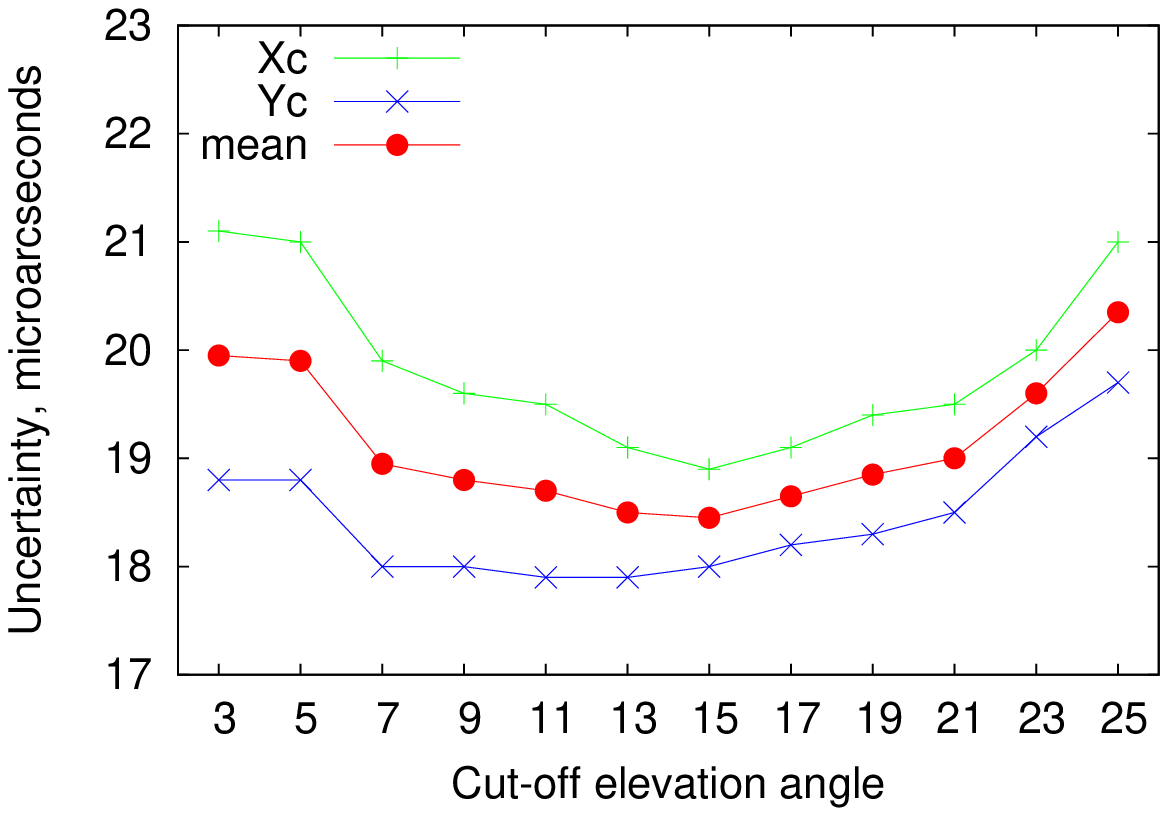}
\hspace{0.02\textwidth}
\epsfclipon \epsfxsize=0.48\textwidth \epsffile{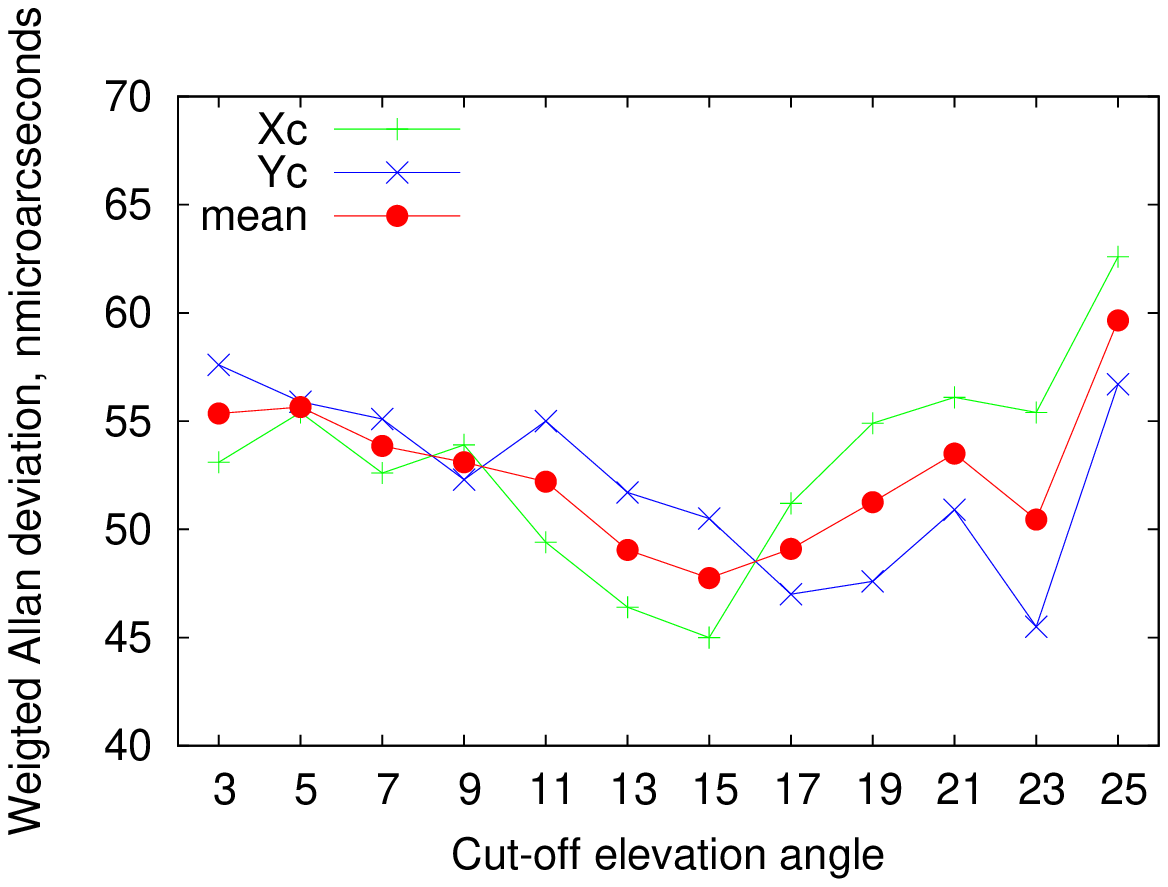} \\
\epsfclipon \epsfxsize=0.48\textwidth \epsffile{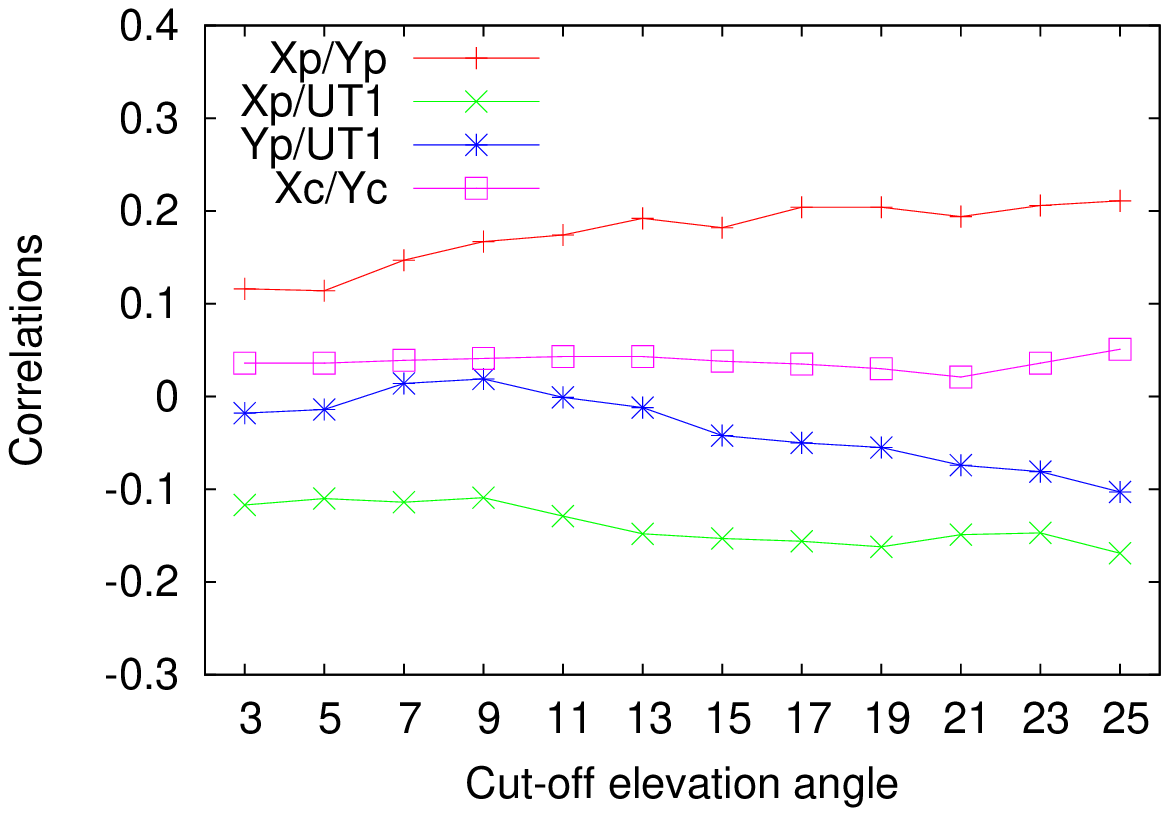}
\caption{Statistics of EOP obtained from CONT05.}
\label{fig:eop_statistics}
\end{figure}

\begin{figure}
\centering
\epsfclipon \epsfxsize=0.48\textwidth \epsffile{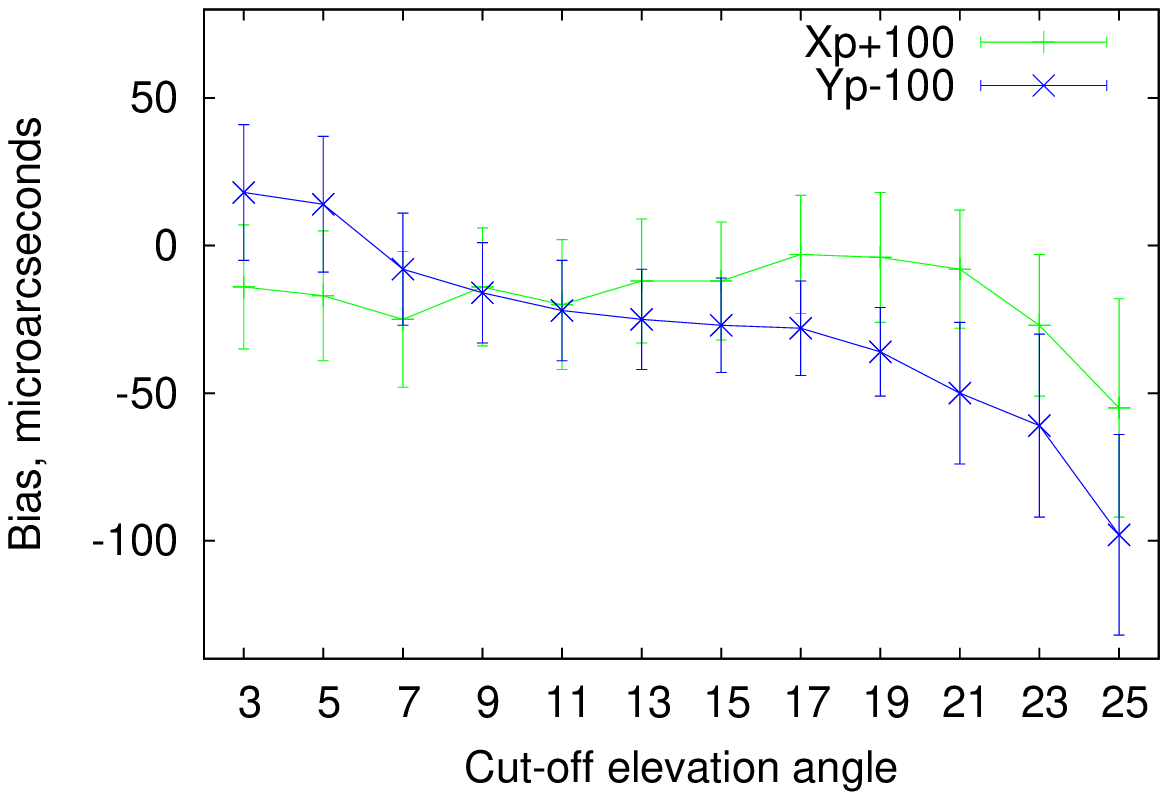}
\hspace{0.02\textwidth}
 \\ 
\epsfclipon \epsfxsize=0.48\textwidth \epsffile{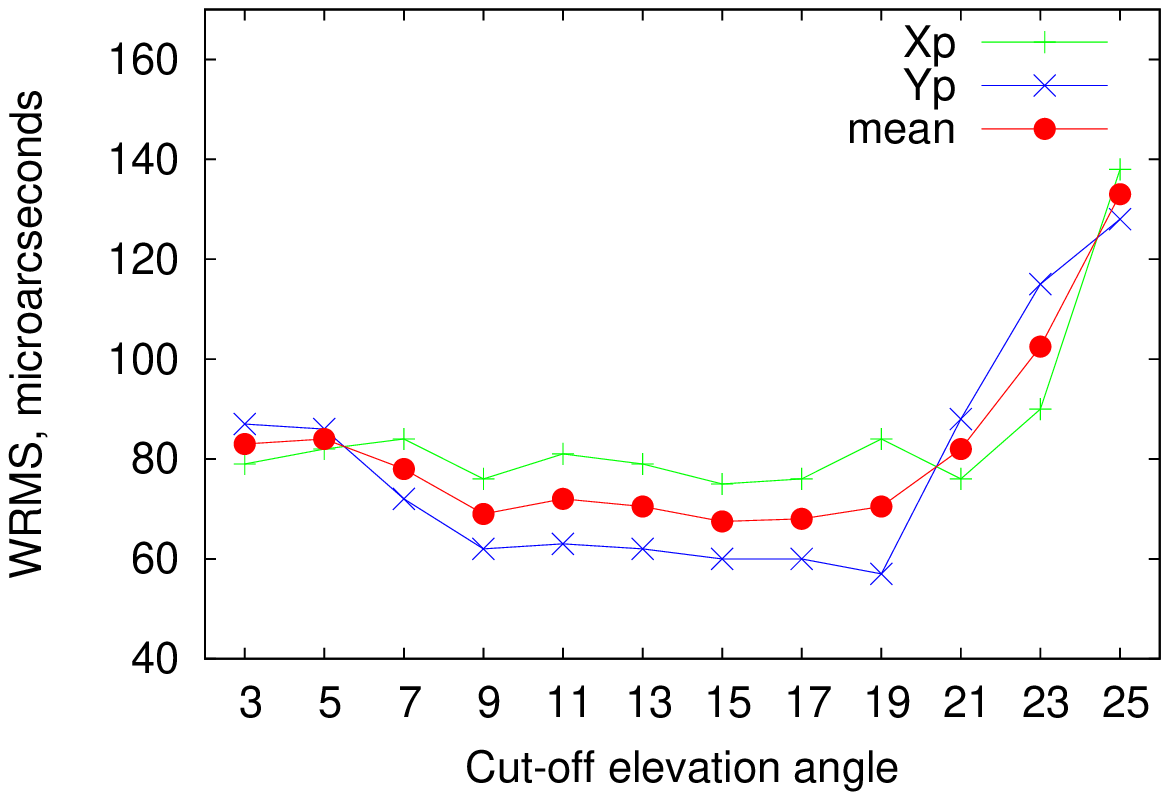}
\caption{Comparison of EOP obtained from CONT05 with the IGS EOP series.}
\label{fig:eop_igs}
\end{figure}

Figure \ref{fig:eop_statistics} shows EOP statistics for different CEA.
Notation used is the following: $Xp$, $Yp$ - terrestrial pole coordinates,
$Xc$, $Yc$ - celestial pole coordinates. Weighted Allan deviation
is computed as described in \cite{Malkin07}.
All the results related to Xp, Yp, Xc and Yc are given
in $\mu$as, the results related to UT1 are given in $\mu$s.
Comparison with IGS time series is shown in Figure~\ref{fig:eop_igs}.
Xp and Yp wrms with respect to IGS EOP series are computed after
removing the bias.

Table \ref{tab:edw_eop} shows the main EDW test results.
Notation of EDW modes is the same as in Table \ref{tab:edw_baseline} with
the 'w\_' prefix omitted. 'All' column correspond to inclusion of
all observations without weighting.
One can see again that several EDW modes show about the same
EOP precision and accuracy.

\begin{table}
\centering
\caption{EOP statistics for different EDW models. See explanation in text.}
\label{tab:edw_eop}
\small          
\tabcolsep=3pt   
\begin{tabular}{|l|c|c|c|c|c|c|c|c|c|}
\hline
\multicolumn{1}{|c|}{Statistics} & All & \multicolumn{8}{|c|}{EDW mode} \\
\cline{3-10}
                    &      & 10\_1& 10\_2& 25\_1& 25\_2& 45\_1& 45\_2& 90\_1& 90\_2 \\
\hline
Xp uncertainty      & 26   & 26   & 25   & 26   & 28   & 27   & 37   & 30   & 51   \\
Yp uncertainty      & 25   & 24   & 24   & 24   & 26   & 25   & 32   & 26   & 43   \\
UT1 uncertainty     & 1.1  & 1.1  & 1.1  & 1.1  & 1.1  & 1.0  & 1.3  & 1.1  & 1.6  \\
Xc uncertainty      & 21   & 21   & 20   & 20   & 19   & 20   & 22   & 22   & 27   \\
Yc uncertainty      & 19   & 18   & 18   & 18   & 18   & 18   & 21   & 20   & 27   \\
\hline
Xp bias w.r.t. IGS  & --79 & --75 & --79 & --77 & --73 & --80 & --86 & --81 & --123\\
Yp bias w.r.t. IGS  & +135 & +135 & +128 & +132 & +129 & +134 & +127 & +129 & +92  \\
\hline
Xp wrms w.r.t. IGS  & 77   & 72   & 76   & 69   & 65   & 68   & 69   & 64   & 86   \\
Yp wrms w.r.t. IGS  & 73   & 74   & 72   & 69   & 61   & 68   & 63   & 71   & 83   \\
mean                & 75   & 73   & 74   & 70   & 63   & 68   & 66   & 68   & 84   \\
\hline
\end{tabular}
\end{table}

\section{Conclusions}

The preliminary conclusions from this test are the following.
\begin{itemize}
\itemsep=-0.3ex
\item The baseline length repeatability steadily grows with the CEA increasing, remaining
  practically the same in the cut-off angle range from 3$^\circ$
 (i.e. no cut-off for the CONT05) to 9$^\circ$.
\item The best result is obtained when the EDW elevation-depending weighting is applied
  to the low-elevation observations.  However, the test results are not always
  unambiguous. Further adjustment of the weighting method may be fruitful.
\item The Xp, Yp and UT1 uncertainties grow with the increasing cut-off angle after
  about 10$^\circ$. Most probably, this reflects the fact that only about 6\% of the
  total number of CONT05 observations were made at the elevations below 10$^\circ$.
  The Xc and Yc uncertainties and scatter depend on the CEA much less.
\item Xp bias w.r.t. IGS slightly depends on the CEA, except the maximum tested CEA values,
  evidently unrealistic.  In contrast, Yp bias substantially changes with increasing CEA.
  Most probably, this can be explained by the CONT05 network orientation, for which the
 longitude of the central meridian $\lambda_0=265^\circ$ just corresponds
 to the Y direction of the terrestrial coordinate system.
\item Some statistics such as the uncertainty and the scatter of the Xc and Yc, as well
 as the WRMS of Xp and Yp w.r.t. IGS have the minimum at the CEA around $15^\circ$,
 which is interesting and deserves a supplement investigation.
\item As one can expect, the correlations between EOP comprising Xp and Yp grow with
 increasing CEA, but remain small due to good CONT05 network geometry.  The same can
 be expected for the IVS2010 network.  The correlation between Xc and Yc remain practically
 the same for all tested CEA, except the maximum tested CEA value, evidently unrealistic.
\end{itemize}

Finally, we can conclude that inclusion of the low-elevation observations,
properly weighted, improves the baseline length repeatability and EOP results.
On the contrary, filtering the observations using the cut-off elevation
method may lead to degradation of geodetic results.
However, this should be mentioned that the conclusions drawn from
the result obtained in this paper has been proven with standard
geodetic VLBI observations, where rather few observations
were made at low elevations, as mentioned above.
Perhaps, special R\&D sessions with more uniform distribution
of observations over the sky, including observations at very low
elevations, may be useful for more detailed study on the impact
and optimal processing of the low-elevation observations on
geodetic parameters obtained from VLBI observations.

It ought be mentioned that all the EDW modes considered in this
paper in fact modify only diagonal elements of the corresponding
covariance matrix. According to Gipson's work \cite{Gipson07} best
result can be achieved in a case of account also for correlations
between observations.
It seems to be interesting to investigate how this approach will
work in KF estimator.

\end{document}